\documentclass[aps,prl,showpacs,twocolumn,floats,superscriptaddress,floatfix]{revtex4-1}
\usepackage{amsmath,amssymb}
\usepackage{graphicx}
\usepackage{psfrag}
\usepackage{graphicx}
\usepackage{epsf}
\usepackage{dcolumn}
\usepackage{bm}
\newcommand{\be}{\begin{eqnarray}}
\newcommand{\en}{\end{eqnarray}}
\newcommand{\ben}{\begin{eqnarray*}}
\newcommand{\enn}{\end{eqnarray*}}
\newcommand{\pa}{\partial}

\newcommand{\f}{\frac}

\newcommand{\ue}{\mathrm{e}}
\newcommand{\ui}{\mathrm{i}}

\newcommand{\bi}{\begin{itemize}}
\newcommand{\ei}{\end{itemize}}

\begin{document}
\title{Nelkin scaling for the Burgers equation and the role of high-precision calculations}
\author{Sagar Chakraborty}
\email{sagar@nbi.dk}
\affiliation{NBIA, Niels Bohr Institute, Blegdamsvej 17, 2100 Copenhagen $\O$, Denmark}
\author{Uriel Frisch}
\email{uriel@oca.eu}
\affiliation{UNS, CNRS, Lab. Cassiop\'ee, OCA, B.P. 4229, 06304 Nice Cedex 4, France}
\author{Walter Pauls}
\email{Walter.Pauls@ds.mpg.de}
\affiliation{Max Planck Institute for Dynamics and Self-Organization, G\"ottingen, Germany}
\author{Samriddhi Sankar Ray}
\email{samriddhisankarray@gmail.com}
\affiliation{UNS, CNRS, Lab. Cassiop\'ee, OCA, B.P. 4229, 06304 Nice Cedex 4, France}
\date{\today}
\begin{abstract}
Nelkin scaling, the scaling of moments of velocity gradients 
in terms of the Reynolds number, is an alternative way of obtaining 
inertial-range information. It is shown numerically and theoretically
for the Burgers equation that this procedure works already for Reynolds 
numbers of the order of 100 (or even lower when combined with a suitable 
extended self-similarity technique). At moderate Reynolds numbers, for the
accurate determination of scaling exponents, it is crucial to use higher than double precision. Similar issues are
likely to arise for three-dimensional Navier--Stokes simulations.
\end{abstract}
\maketitle
\noindent Nelkin \cite{N90}, showed that the multifractal model of turbulence
\cite{M74,PF83}, implies certain scaling relations for moments of velocity
gradients (henceforth \textit{gradmoments}).  According to Nelkin, at high Reynolds numbers, when plotted as a
function of the Reynolds number $R$, the $p$th moment of any component $\nabla
u$ of the velocity gradient should scale, to leading order, as  \be
\langle (\nabla u)^p \rangle \sim R ^{\chi_p}.  \en The exponents $\chi_p$ are
expressible in terms of the multifractal structure function exponents
$\zeta_p$ (cf. \cite{N90} or \cite{Fbook}, Sec. 8.5.6).

By using very highly resolved direct numerical simulation, it has been
checked by Schumacher, Sreenivasan and Yakhot that not only is such
scaling present (its first verification), but that it is already seen at Reynolds numbers
around 200, well below those where structure functions show any
inertial-range scaling \cite{SSY07}. This is perhaps not so suprising,
given that inertial-range scaling is for intermediate asymptotics with
\textit{two} large parameters, the Reynolds number and the ratio of
the scale under consideration to the typical dissipation scale,
whereas Nelkin scaling just requires a large Reynolds number.

The one-dimensional Burgers equation
\begin{equation}
\partial_t u +u\partial_x u = \nu \partial_x^2 u; \quad u(x,0) =u_0(x),
\label{burgerseq}
\end{equation}
where $u$ is the velocity and $\nu$ the kinematic viscosity, can
can throw light on why gradmoments display good scaling at such moderate
Reynolds numbers. Furthermore, it allows analytical determination of
all the dominant and subdominant terms in the high-Reynolds number
expansion of gradmoments.
We note that in a recent paper \cite{CFR10}, the Burgers equation was used to
illustrate why the extended self-similarity (ESS) technique \cite{BCTBMS93}
gives improved scaling through the depletion of subdominant corrections.

Heuristically, it is quite simple to show that for the Burgers equation we
expect $\chi_p = p-1$. Indeed, at high Reynolds numbers, the solutions of
\eqref{burgerseq} display shocks broadened by viscosity over a distance
$O(\nu) = O\left(R^{-1}\right)$. Within a shock, the $p$th
power of the velocity gradient is $O\left(R^{p}\right)$. Since
shocks cover a fraction $O\left(R^{-1}\right)$ of the
one-dimensional spatial domain, the stated scaling results. Of course such an
argument tells us nothing about subdominant corrections and thus cannot be
used to predict at what kind of Reynolds numbers this scaling emerges.  

We shall now address these issues more systematically, using simulations and
theory. We shall also address a new question: scaling exponents 
are notoriously known with poor accuracy (cf., e.g., \cite{Fbook});
how accurately can we determine such exponents by working with Reynolds number at which there are
significant subdominant corrections to scaling? Using recent results of van
der Hoeven \cite{V09, PF07}, we shall show that this requires a subtle
tradeoff between Reynolds numbers and   precision (number of decimal digits)
used in the calculations.  
\\

We begin with simulation-based results for the Reynolds number
dependence of gradmoments when standard double-precision
calculations suffice. We follow here the same strategy as
in Ref.~\cite{CFR10}: we solve the Burgers equation \eqref{burgerseq}
with the initial condition $u_0(x) = \sin x$, using a pseudo-spectral method
combined with fixed-time-step fourth-order Runge--Kutta time marching and a slaved scheme,
known by the acronym ETDRK4 \cite{CM02}, for handling the viscous 
dissipation. The gradmoments of integer order $p$, as a function of the Reynolds
number
$R \equiv 1/\nu$, are defined
as spatial averages over the period $2\pi$:
\begin{equation}
M_p(R)\equiv \f{1}{2\pi}\int_0 ^{2\pi} dx\, \left[\f{\pa u(x,t)}{\pa x}\right]^p.
\label{deterMp}
\end{equation}
%
%
%
%
Gradmoments are calculated for orders $p$ from two to ten and Reynolds numbers
$R$ from twenty to one thousand. The number of collocation points $N$ is taken
between 8K and 256K where K stands for $2^{10}= 1024$; the time step $\delta
t$ is between $10^{-5}$ and $10^{-6}$. We checked that the errors on
gradmoments stemming from spatial and temporal truncation stay below the
level needed for a double-precision calculation. The output is calculated at
$t=2$ when the solution of the Burgers equation has a well-developed
shock. Since, as explained above, gradmoments are expected to behave
asymptotically as $R^{p-1}$ at large $R$, we display them in
\textit{compensated} manner, that is divide them by
$R^{p-1}$. 
\begin{figure}
  \includegraphics[height=6cm,width=9cm]{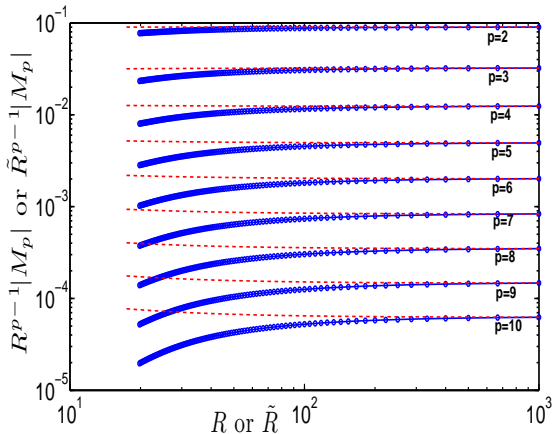}
  \caption{(Colour online) Compensated $p$th-order moments ($p$ from 2 to 10) of velocity 
gradient ($M_p$) versus both $R$ (continuous line with points in blue) and the ESS-type surrogate
$\tilde{R}$ (dashed red line).}
\label{f:compgradmom}
 \end{figure}
Figure \ref{f:compgradmom} shows the compensated gradmoments as a
function of Reynolds number. Visual inspection shows that the expected flat
behavior of the compensated gradmoments sets in around $R =40$ for the lowest
order $p =2$ and around $R= 300$ for $p=10$. In contrast, inertial-range
scaling for structure functions, calculated from the same solution of the
Burgers equation, appears clean only around Reynolds numbers of several
thousands \cite{CFR10}. This discrepancy, of nearly two orders of magnitude,
can be made even larger by resorting to a procedure inspired from ESS in which one resorts to a surrogate
of the spatial separation, such as the third-order structure function and
plots structure functions versus the surrogate. In the case of gradmoments, we
observe that the mean energy dissipation is given in terms of the mean square
velocity gradient by $\varepsilon=\nu M_2 = (1/R) M_2$. This has a
finite positive limit $\varepsilon_\infty$ as the Reynolds number
tends to infinity. Hence, we can use
$\tilde R \equiv M_2/\varepsilon_\infty$ as a (suitably normalized) surrogate of the Reynolds number.
This we call ESS-type plotting. The same Fig.~\ref{f:compgradmom} also
shows this type of plotting.  Now, the data look almost completely
flat, except for the largest value of $p$ around $ \tilde R =20$ where the
data bend slightly upwards, as revealed by looking at the figure from
the side \footnote{This ESS technique can be readily extended to 
three-dimensional Navier--Stokes experiments and simulations.}.

Of course, all this has to do with subdominant corrections to scaling
and the way they are affected by the ESS-type procedure. We now turn
to theoretical interpretations.  For this we use the exact solution of
the Burgers equation, obtained by
employing the Hopf--Cole method \cite{H50,C51} that transforms the Burgers
equation into the heat equation. For the initial condition $u(x,0)=\sin(x)$ 
this solution reads
\begin{eqnarray}
u(x,t)&=& -2 \nu \partial_x\ln \theta(x,t) \label{hopfcolesoln}\\
\theta(x,t) &=& \int_0^{2\pi}  \ue ^{\cos (x-x')/(2\nu)}\,
G(x',t) \,dx' \label{heat}.
\end{eqnarray}
Here, $G(x',t)= \sum_{k= -\infty}^{k=\infty} \ue ^{\ui kx'-\nu
k^2t}$ is the Green's function for the heat equation in the $2\pi$- periodic 
case. 
We want to use this solution to determine the asymptotics of gradmoments for
small $\nu$, i.e., large $R$.  
Using the method of steepest descent, in a way somewhat similar to what is
found in Ref.~\cite{scooped}, one can show that, for large $R$ and any integer $p\ge 2$ 
\begin{equation}  
  M_p(R) = A_p R^{p-1} + B_pR^{p-2}+C_pR^{p-3} + \ldots
\label{mpexpansion}
\end{equation}
The coefficients are given  by rather complicated and numerically
ill-conditionned integrals. 

The expansion \eqref{mpexpansion} and the numerical values of the
coefficients can actually be obtained by an alternative 
semi-numerical procedure, called 
\textit{asymptotic extrapolation}, developed by van der
Hoeven~\cite{V09} (see also \cite{PF07} for an elementary presentation). Let us now say a few
  words
about this technique, which will also be used below in connection with 
high-precision spectral calculations. Suppose we have determined
numerically with high precision the values of a function $f(n)$ for
integers $n$ up to some high value $N$. We wish to obtain from this
as many terms as possible in the high-$n$ asymptotic expansion of $f$.
Trying to fit the function by a guessed leading asymptotic form with some
free parameters, will generally lead to very poor accuracy in
such parameters. With some information about the structure
of the various terms in the expansion, a better method is to fit
an expression containing one or several subdominant corrections (all
with some unknown parameters). Lacking such information, asymptotic extrapolation handles the
problem by applying to the data a 
sequence of suitably chosen transformations  that successively 
strip off the dominant and subdominant terms in the expansion 
for large $n$. At certain stages of such transformations, 
the processed data allow simple extrapolations, most often 
by a constant. The transformations are meaningful as long as the 
successively transformed data is free from conspicuous rounding noise and $n$ has reached a simple asymptotic behavior (e.g. flat). From the extrapolation stages, it 
then becomes possible (by undoing the transformations made) to 
obtain the asymptotic expansion of the data (including the values of
the various parameters) up to some order which depends on the
precision 
of the data and on the value of $N$.  
Here, we will denote the transformations by using the notation
of Ref.~\cite{PF07}. Thus, {\bf I} stands for ``inverse'', {\bf R} for ``ratio'',
{\bf SR} for ``second ratio'' and {\bf D} for ``difference''. The sequence of transformations is chosen 
through various tests which  provide some clue about the asymptotic class 
in which the data falls.
 
\begin{table*}
\framebox{\begin{tabular}{l|l|l|l|l|l|l}
order$(p)$ & $\chi_p$ & $A_p$ & $\chi^{(1)}_p$ & $B_p$ & $\chi^{(2)}_p$ & $C_p$ \\
\hline
 2 & 0.9999987 & +0.09032605 & -- 0.002 & -- 0.2290236 & -- 1.002 & +0.2011 \\ 
 3 & 1.999998 & -- 0.03245271 & 1.00001 & +0.1736854 & 0.005 & -- 0.1325\\
 4 & 2.999996 &  +0.01249279 & 2.00001  & -- 0.090466 & 1.0001 & +0.08417 \\
 5 & 3.999995 & -- 0.00498725 & 3.00001  &  +0.045622 & 1.99988 & -- 0.08209\\
 6 & 4.999994 &  +0.00203621 & 4.00001  & -- 0.022523 & 2.99993 & +0.06103\\
 7 & 5.999993 & -- 0.00084414 & 5.000008  & +0.010955 & 4.0002 & -- 0.0398\\
 8 & 6.999992 &  +0.0003539 &  5.999993 & -- 0.00526 & 5.002 & +0.024\\
 9 & 7.999994 & -- 0.0001495 & 6.99991 &    +0.0025  & 6.009 & -- 0.01\\
 10 & 9.00001 &  +0.000063 &  7.9995  & -- 0.0012 & 7.03 & +0.03 \\
\end{tabular}}

\caption{Dominant scaling exponents $\chi_p$ and the first two subdominant
  exponents $\chi_p^{(1)}$ and $\chi_p^{(2)}$ together with the 
corresponding coefficients $A_p$, $B_p$, and $C_p$ for the large-$R$ 
behavior of gradmoments of order $p$, obtained by asymptotic extrapolation processing of a 400-digit precision determination of gradmoments from the Hopf-Cole solution. The theoretical values are 
$\chi_p = p - 1$, $\chi_p^{(1)} = p - 2$, and $\chi_p^{(2)} = p - 3$.}   
\label{t:hopfcoleexpansion}
\end{table*}
To apply asymptotic extrapolation to the determination of the coefficients in the
high-Reynolds number  expansion \eqref{mpexpansion}, we calculate the
Hopf--Cole solution \eqref{hopfcolesoln}-\eqref{heat} and the
gradmoments \eqref{deterMp} using extreme precision floating point 
calculations \cite{mpfun}  with
400 decimal digits. This precision guarantees that
the only source of errors is lack of simple asymptoticity. The convolution structure of \eqref{heat}
allows the use of fast Fourier transforms, also in very high precision \cite{fft}, for calculating $\theta$, $u$ and  various space derivatives. The
 Reynolds number $R$ is given all integer values from $18$
to $R_{\rm max}= 400$. The processing of the gradmoments for $p$ from 2 to 10
involves typically 15 stages of transformations, the first eight of which
are always {\bf R - 1}, {\bf I}, {\bf D}, {\bf D}, {\bf I}, {\bf D},
{\bf D}, {\bf D} \footnote{{\bf R - 1} means applying {\bf R} and then
  subtracting unity.}. From the undoing of the transformations, using the ``most asymptotic data
points'' for determining constants, we obtain the following
expansion:
\begin{equation}  
  M_p(R) = A_p R^{\chi_p} + B_pR^{\chi_P^{(1)}}+C_pR^{\chi_P^{(2)}} + \ldots
\label{tsmpexpansion}
\end{equation}
The results are shown in Table~\ref{t:hopfcoleexpansion}. Only those digits of the coefficients
that agree when processing the data successively with $R_{\rm max}= 200$
and  $R_{\rm max}= 400$ are shown. It is seen that the scaling
exponents for the dominant term $\chi_p$ and the first and second
subdominant terms, $\chi_P^{(1)}$ and $\chi_P^{(2)}$ are very close
to their theoretical values obtained  from \eqref{mpexpansion}. The relative 
discrepancies are in the range $10^{-5}$ -- $10^{-6}$ for the dominant exponent
and the accuracy degrades for subdominant corrections, as expected. 

From the expansion \eqref{tsmpexpansion} we can readily understand why
Nelkin scaling appears at rather moderate Reynolds number: the absolute
value of relative
correction stemming from the first subdominant term is $R^{-1}|B_p/A_p|$. 
For example, it reaches the ten percent level which is easily picked
up visually at $R = 10 |B_p/A_p|$. Table~\ref{t:howlarge} shows
the values of $|B_p/A_p|$ and we now understand why  flat compensated
gradmoments are seen in Fig.~\ref{f:compgradmom} beyond Reynolds
numbers, varying with $p$, from a few tens to a few hundreds. To understand
the ESS-type even better scaling, we expand the surrogate $\tilde R$ in
terms of  $R$. From \eqref{mpexpansion} with $p=2$ and noticing that
$A_2 =\varepsilon_\infty$,  we obtain
\begin{equation}  
\tilde R = R^1+\frac{B_2}{A_2} R^0+ O\left(R^{-1}\right).  
\label{rrtilde}
\end{equation}
Eliminating $R$ between \eqref{mpexpansion} and \eqref{rrtilde}, we obtain
\begin{equation}  
M_p = A_p {\tilde R}^{p-1}+  \tilde B_p  {\tilde R}^{p-2} +\ldots; \,
{\tilde B_p} = B_p-\frac{(p-1)A_p}{A_2}B_2.
\label{mpdeputyexpansion}
\end{equation}
Note that the expansion in terms of the surrogate $\tilde R$ has the same
structure as \eqref{mpexpansion} and precisely the same dominant-term
coefficient $A_p$. However the coefficient $\tilde B_p$ of the first
subdominant correction is significantly smaller than $B_p$ (in absolute value)
and may have a different sign.  This explains for example why the graph for
the compensated third-order gradmoment in terms of $R$ bends down at the low
end while it bends very slightly up in terms of $\tilde R$. As a consequence
of the reduced subdominant corrections, the asymptotic behavior of gradmoments
in the ESS-type representation emerges at Reynolds numbers 5 to 20 times
smaller than in the ordinary representation (see Table~\ref{t:howlarge}).\\
\begin{table}[h]
\framebox{\begin{tabular}{c|c|c}
order$(p)$ & $R_p^\star = |B_p/A_p|$&$\tilde R_p^\star= |\tilde B_p/A_p|$\\
\hline
 2 & 2.5344  & 0.0\\ 
 3 & 5.3520  & 0.2827 \\
 4 & 7.2414 & 0.3622 \\
 5 & 9.1477 & 0.9906 \\
 6 & 11.0613 & 1.6116 \\
 7 & 12.9785 & 2.2290 \\
 8 & 14.8980 & 2.8440 \\
 9 & 16.8222  & 3.4544 \\
 10 & 19.0604 & 3.7507 \\
\end{tabular}}
\caption{Estimates of Reynolds numbers beyond which subdominant corrections
  become small in the Reynolds number representation (middle column) and in the ESS-type representation (last column).}
\label{t:howlarge} 
\end{table}
We should not be carried away and state that good scaling can emerge already
at very moderate Reynolds number provided we take the right quantity (here,
gradmoments) and the right data processing technique (here, ESS). It all
depends on what we call ``good scaling''. If we want to obtain scaling
exponents with an error not exceeding $10^{-2}$ or $10^{-3}$, a \textit{flat
  looking} compensated graph is definitely not enough since this is achieved
as soon as the relative error is somewhere below $10^{-1}$. We now address the
issue how asymptotic (how large in Reynolds number) and how precise should a
spectral calculation be in order to truly give accurate scaling exponents. Of
course, the higher the Reynolds number, the lower the relative subdominant
corrections will be.  But, without enough precision, the simultaneous
determination of dominant terms and subdominant corrections, say by asymptotic
extrapolation, will be unable to handle more than very few such corrections
and thus gives us substantial errors in the final results.  In order to be
closer to more realistic models such as the multi-dimensional Navier--Stokes
equations, in investigating the trade-off between asymptoticity and precision,
we refrain from using the exact solution of the Burgers equation and resort to
time integration by (pseudo-)spectral technique.  We use double and quadruple
precision, both combined with asymptotic extrapolation, so as to obtain the
most accurate possible parameters. We calculate the scaling exponents
$\chi_4$ and $\chi_6$ of the fourth and the sixth gradmoments, whose theoretical
exact values are three and five, respectively. We determine how accurately we can predict these exponents when applying
asymptotic extrapolation (which for this purpose is substantially better than
the aforementioned ESS technique), using various maximum Reynolds numbers
$R_{\rm max}$. In double precision we were able to use three stages 
and in quadruple precision eight stages of the aforementioned transformations.
 \begin{figure}
  \includegraphics[height=8cm,width=9cm]{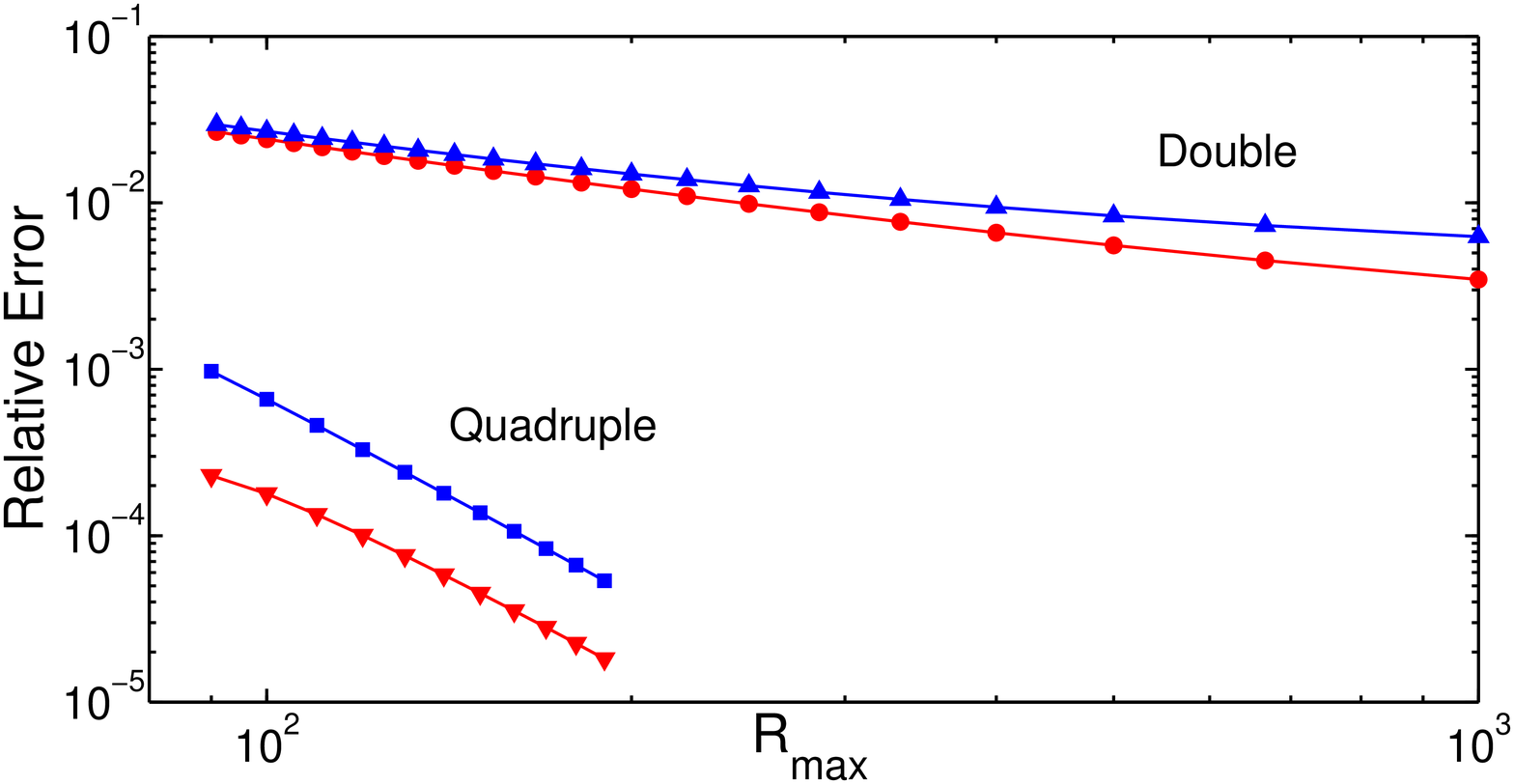}
  \caption{(Color online) Relative error of  Nelkin exponents $\chi_4$ and
    $\chi_6$ obtained by asymptotic extrapolation from 
  pseudo-spectral calculations up to a maximum Reynolds number $R_\textrm{max}$. Upper
  set of curves: double precision calculations ($\chi_4$: red filled circles, 
  $\chi_6$: blue filled triangles); lower set of curves: quadruple
precision ($\chi_4$: red inverted triangles, $\chi_6$:
  blue filled squares).}
\label{f:percenterror}
 \end{figure}
The maximum wavenumber and the size of the time step are
the same as reported at the beginning of the paper. We checked, by further 
halving of spatial and temporal resolutions, that they contribute negligible
errors to the result. Figure~\ref{f:percenterror} shows the relative errors for the
two  types of precision as function of $R_{\rm max}$. It is striking that, when 
doubling the precision we can decrease the Reynolds number by about a
factor of eleven (from 1000 to 90) and still obtain a substantial decrease (by a factor 
of 3 to 10) in the relative error.
For accurate determination of scaling exponents, increasing the
precision is here definitely more efficient than increasing the Reynolds number.
It remains to be seen if this result carries over to a much broader
class of equations, including multi-dimensional incompressible
problems  displaying random behavior. Already, we can state that the
use of Nelkin scaling to analyze multifractal scaling  in 
simulated 3D turbulent flow should definitely be encouraged, and
preferably combined with high precision caculations.

%
%
%
\begin{acknowledgements}
We are indebted to Joris van der Hooven, T. Matsumuto, D. Mitra,  O. Podvigina, and V. Zheligovsky for a number of useful discussions.
S.C. thanks academic and financial support rendered by NBIA (Copenhagen); and Danish Research Council for a FNU Grant No. 505100-50 - 30,168. The work was partially supported by
ANR ``OTARIE'' BLAN07-2\_183172.  Some of the computations used the M\'esocentre de calcul
of the Observatoire de la C\^ote d'Azur. 
\end{acknowledgements}


\begin{thebibliography}{99}
%
\bibitem{N90} M. Nelkin, Phys. Rev. A {\bf 42}, 7226 (1990).
\bibitem{M74}B. Mandelbrot, {J. Fluid Mech. {\bf 62}}, 331 (1974).
\bibitem{PF83} G. Parisi and U. Frisch, in {\it Turbulence and Predictability in Geophysical Fluid Dynamics and Climate Dynamics}, edited by M. Ghil, R. Benzi and G. Parisi (North-Holland, Amsterdam, 1985), p. 84.
\bibitem{Fbook}U. Frisch, {\it Turbulence --- The Legacy of A. N. Kolmogorov}, (Cambridge University Press, Cambridge, 1995).
\bibitem{SSY07}J. Schumacher, K. R. Sreenivasan and V. Yakhot, New J. Phys. {\bf 9}, 89 (2007).
\bibitem{CFR10} S. Chakraborty, U. Frisch and S. S. Ray, {J. Fluid Mech. {\bf {649}}}, 275 (2010).
\bibitem{BCTBMS93}R. Benzi, S. Ciliberto, R. Tripiccione, C. Baudet, F. Massaioli and S. Succi, Phys. Rev. E \textbf{48}, R29 (1993).
\bibitem{V09} J. van der Hoeven, {J. Symb. Comput.} {\bf 44}, 1000 (2009).
\bibitem{PF07} W. Pauls, and U. Frisch, {J. Stat. Phys.} {\bf 127}, 1095 (2007).
\bibitem{CM02} S. M Cox and P. C. Matthews, {J. Comp. Phys.} {\bf 176}, 430 (2002).
\bibitem{H50} E. Hopf, {Comm. Pure Appl. Math.} {\bf 3}, 201 (1950).
\bibitem{C51}J. D. Cole, {Quart. Appl. Math.} {\bf 9}, 225 (1951).
\bibitem{scooped} D. O. Crighton and J. F. Scott, { Phil. Trans. Roy. Soc. London \bf{292}}, 101 (1979).
\bibitem{mpfun} D. H. Bailey, ``High-Precision Arithmetic in Scientific Computation'', 
Computing in Science and Engineering, May-June, 2005, pg. 54-61; LBNL-57487.
See also http:$//$crd.lbl.gov$/\sim$dhbailey$/$
\bibitem{fft} http:$//$www.kurims.kyoto-u.ac.jp$/\sim$ooura$/$fft.html.

 %
\end{thebibliography}
\end{document}